\documentclass[%
 reprint,
 amsmath,amssymb,
 aps,
]{revtex4-1}
\usepackage{graphicx}
\usepackage{dcolumn}
\usepackage{bm}
\usepackage{comment}
\usepackage{upgreek}
\usepackage{todonotes}
\usepackage{xcolor}

\begin{document}
\preprint{APS/123-QED}

\title{Topological defects in buckled colloidal monolayers}

\author{Aaron L. Galper}
 \thanks{These authors contributed equally}
 \affiliation{%
 Department of Physics, Harvey Mudd College, Claremont, CA
}%

\author{Henrik N. Barck}%
 \thanks{These authors contributed equally}
 \affiliation{%
 Department of Physics, Harvey Mudd College, Claremont, CA
}%

\author{Conor M. Floyd}%
 \affiliation{%
 Department of Physics, Harvey Mudd College, Claremont, CA
}%

\author{Elliot A. Snyder}%
 \affiliation{%
 Department of Physics, Harvey Mudd College, Claremont, CA
}%

\author{Charlie J. Schofield}%
 \affiliation{%
 Department of Physics, Harvey Mudd College, Claremont, CA
}%

\author{Sorin A. P. Jayaweera}%
 \affiliation{%
 Department of Physics, Harvey Mudd College, Claremont, CA
}%

\author{Ian G. McGuire}%
 \affiliation{%
 Department of Physics, Harvey Mudd College, Claremont, CA
}%

\author{Sharon J. Gerbode}
 \email{gerbode@hmc.edu}
\affiliation{%
 Department of Physics, Harvey Mudd College, Claremont, CA
}%

\date{\today}

\begin{abstract}
When colloidal particles are vertically confined to a gap of between $\sim \! 1.3$-$1.6$ particle diameters, they pack into buckled crystals of particles in either ``up'' or ``down'' states. Neighboring particles tend to occupy opposite states, analogous to the behavior of antiferromagnetic spins. The particles sit on a nearly-triangular lattice, and the spins of trios of adjacent particles are geometrically frustrated. Two levels of translational order exist in this system: that of the underlying triangular lattice in the horizontal plane, and that of the emergent frustrated spin lattice in the vertical dimension. We study the topological defects of both levels of translational order, and we find that both types of defects play a role in crystal grain boundary structure and spin domain coarsening. We classify the spin defects and outline the basic rules for their motion, and we observe interactions between dislocations and spin defects. Finally, we map the phase space of spin coarsening in the buckled monolayer, characterizing which types of defects drive the dynamics. Understanding defect formation, motion, and interaction in the buckled monolayer is the first step in predicting the material properties and aging of this geometrically frustrated, self-assembled system.
\end{abstract}


\maketitle

\section{Introduction}
If crystalline materials did not contain defects, they could not be deformed, shaped or customized for applications. Understanding crystal defects is crucial to the performance of devices made from crystalline materials; the distribution and motion of defects within crystalline materials determine properties ranging from yield strength to electrical conductivity \cite{Meyers2006,Kreuer2003,Huang2011,HirthLothe1982}. The experimental discovery of grain boundaries and dislocations, and the subsequent development of the theory of dislocations in the 20th century, closed holes in our understanding of how crystalline materials deform under stress and how crystalline domains coarsen with age \cite{Burgers1940,Read1950,Cottrell1949,HirthLothe1982}. 

Experimental studies of defects in colloidal crystals have illuminated a number of fundamental processes such as dislocation glide \cite{Schall2004, Gerbode2008}, grain rotation \cite{Irvine2013,Lavergne2018, BarthMartinez2021}, work hardening \cite{Kim2024}, and the motion of grain boundaries \cite{Skinner2010, Gokhale2013, Chaffee2024}. Studies of colloidal crystals have provided elegant toy models for investigating statistical mechanical processes that also occur in other crystalline systems such as metals.

\begin{figure}[h!]
 \includegraphics[width=0.5\textwidth]{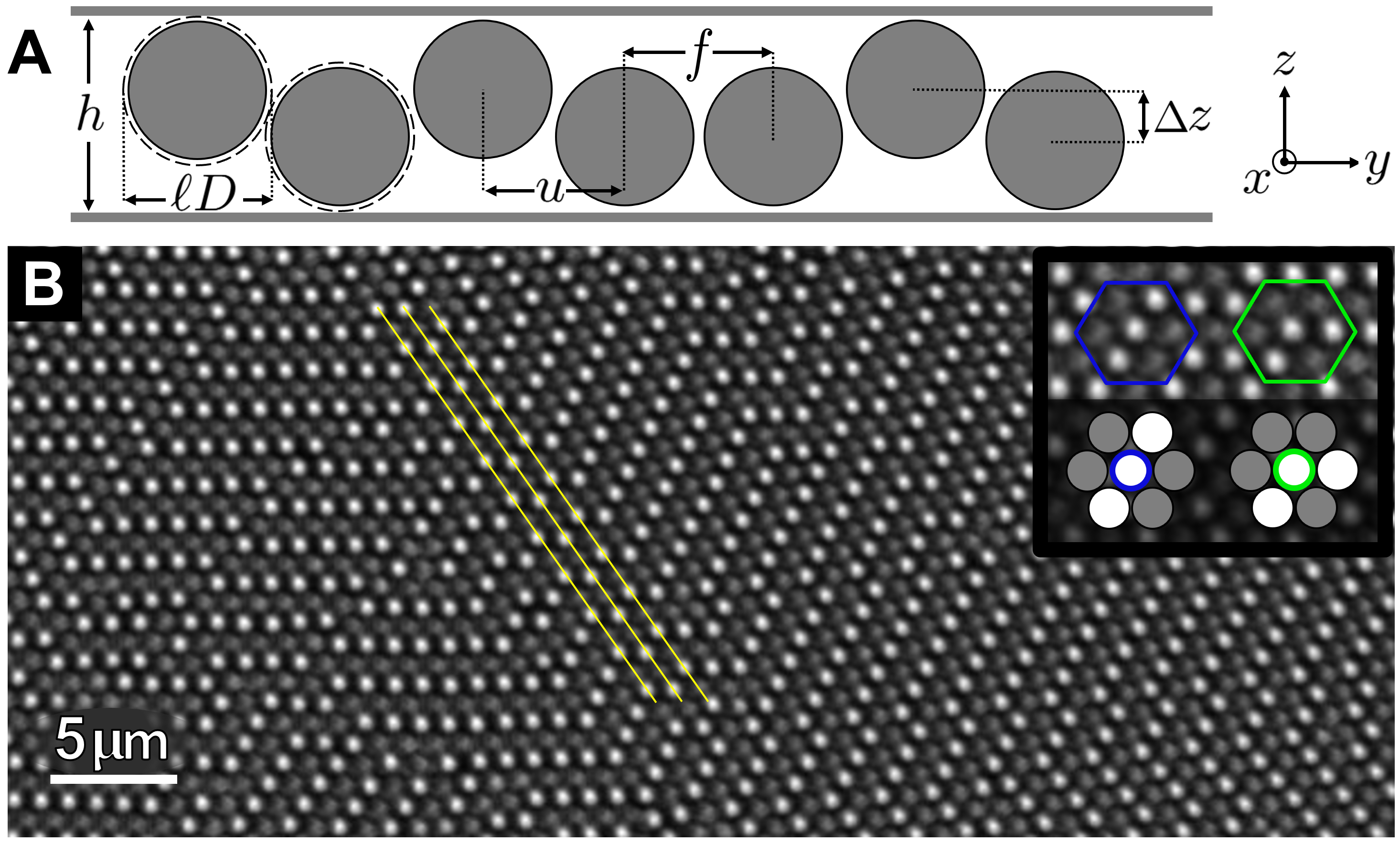}
  \caption{Frustrated spin order in a buckled colloidal monolayer. (a)~Simplified side view schematic of sample with gap height $h = \ell D+\Delta z$ in the $z$ direction, where $D$ is the particle diameter, $\ell$ is the dimensionless looseness parameter, and $\Delta z$ is the ``free height'' between particle centers of opposite spin particles. In the schematic, the centers of same-spin neighbors are separated by distance $f=\ell D$ along the $y$ direction, while the centers of opposite-spin neighbors are spaced by only $u=\sqrt{\ell^2D^2-(h-\ell D)^2}$. (b) A buckled monolayer of colloidal particles. Particles pack into one of many ground state domain configurations, consisting of stripe and zig-zag motifs (highlighted in inset). Ground state domains are composed of rows of alternating spins along a single axis, such as the three rows highlighted with yellow lines.} 
  \label{fig:1}
\end{figure}

The presence of geometric frustration, for example in the classic case of the antiferromagnetic Ising model on a triangular lattice \cite{Wannier1950, Wannier1973} can influence crystal strain \cite{Hayashi2006, Lee2000, Chen1986, Gu1996} and therefore the presence and motions of crystal defects. The buckled monolayer is a geometrically frustrated colloidal system that has been previously studied experimentally in colloidal crystals and macroscale models \cite{Murray1994,Han2008, Zhou2017, Wang2025} and using theoretical and  computational models \cite{Shokef2009, Shokef2011, Dublenych2013}. In our buckled monolayer system, a single layer of hard sphere colloidal particles form a nearly triangular lattice in the $x$-$y$ plane, but each particle has the additional freedom to move slightly up or down in the $z$ direction, bounded by two glass plates separated by a gap of height $h \sim 1.3$-$1.6$ particle diameters. A side view schematic of a single row of particles within the buckled monolayer is shown in Fig.~\ref{fig:1}(a). To pack as efficiently as possible and maximize the entropy, adjacent particles buckle up or down. The ground state configuration for a single line of such particles alternates up and down like spins in a 1D antiferromagnet. By analogy to an antiferromagnetic Ising model, we can identify this alternating pattern as the ``spin order" and we call the up/down states ``spin." In a 2D triangular lattice, a particle cannot have opposite spins to all of its neighbors. Consequently, the spin order is geometrically frustrated. Since these frustrated pair interactions cannot be satisfied everywhere on a triangular lattice, there is no single ground state. Instead, particles pack into one of many degenerate ground states, which are composed of parallel lattice lines of alternating spins, as shown in Fig.~\ref{fig:1}(b). 

We find that the frustrated order influences the motion of $x$-$y$ lattice dislocations (henceforth called ``lattice dislocations'') within the buckled monolayer. Notably, the alternating spin order of the ground state causes an anisotropic compressibility of the triangular lattice lines, which impacts lattice dislocation glide. We also identify a new set of topological defects called ``spin defects.'' These translational defects of the frustrated order of the $z$ coordinates have cores of two adjacent same-spin particles on an otherwise alternating row. Using both experiments and simulations, we find that spin defects form and move as point defects or dislocations on a sublattice of only opposite-spin nearest neighbor edges. 

Both lattice dislocations and spin defects coarsen regions of ordered spins. We present a phase diagram which indicates where each type of topological defect plays a dominant role in spin coarsening. Understanding these defect processes is the first step in modeling the material properties and time evolution of this frustrated crystalline material, and may serve as a guide to understanding defects in other geometrically frustrated systems.

\section{Buckled colloidal monolayers experimental methods}
We prepare buckled monolayer colloidal polycrystals using either silica microspheres (diameter $D\sim1.3~\upmu$m, Sekisui Micropearl Spacers, Dana Enterprises International, CA) or polystyrene microspheres (diameter $D\sim1~\upmu$m, Duke Scientific, Palo Alto CA) suspended in water. We confine the colloidal suspension in a sealed wedge-shaped glass cell as previously described \cite{Gerbode2008}, and tilt the cell to allow the particles to sediment into the gap. The gap height $h$ varies along the length of the cell, filtering out any larger particle clusters, yielding a monodisperse flat monolayer crystal at $h=D$ and a buckled monolayer where $h \in [1.3 D, 1.6D]$. Larger gap heights yield square packing crystal structures, and then buckled multilayers as previously described \cite{Murray1994}. In the buckled monolayer region, the particles pack with average center-to-center separation $\ell D$, where $D$ is the particle diameter and $\ell$ is a dimensionless looseness parameter that varies between $\sim\!1.03$-$1.08$ in our experiments. 

We observe the buckled monolayer using an inverted light microscope. Because of the wedge-shaped cell, the buckled monolayer is imperfectly aligned with the focal plane; we correct this misalignment using a custom adjustable microscope stage plate, as described in supplemental section S1 \cite{suppmat}. We have developed image processing techniques that expand on standard featuring algorithms \cite{Crocker1996} to locate both bright and dark particles in the buckled monolayer and sort particles as either buckled ``up'' or ``down'', as described in supplemental section S2 \cite{suppmat}.

We identify local spin patterns, or motifs, as previously defined by Han et al \cite{Han2008}. In this study, our primary focus is on the two motifs that when appropriately tiled, allow for optimal particle packing in the $x$-$y$ plane: the stripe and the zig-zag, highlighted in the inset of Fig.~\ref{fig:1}(b). These stripe and zig-zag motifs are therefore the only motifs that can be present in the ground state of the system, so they are ``ground state'' motifs. We call regions of optimally packed particles consisting of only ground state motifs ``spin domains''.

\section{Lattice dislocations}
\label{dislocations}
Buckled monolayer crystals have two levels of order: the translational order of the nearly-triangular lattice formed by particle $x$-$y$ coordinates, and the spin order of their $z$-coordinates. The topological defect of the translational $x$-$y$ order is the lattice dislocation, which in a two dimensional triangular lattice has at its core a pair of particles, one with five nearest neighbors and one with seven. 

\begin{figure}[h!]
 \includegraphics[width=0.48\textwidth]{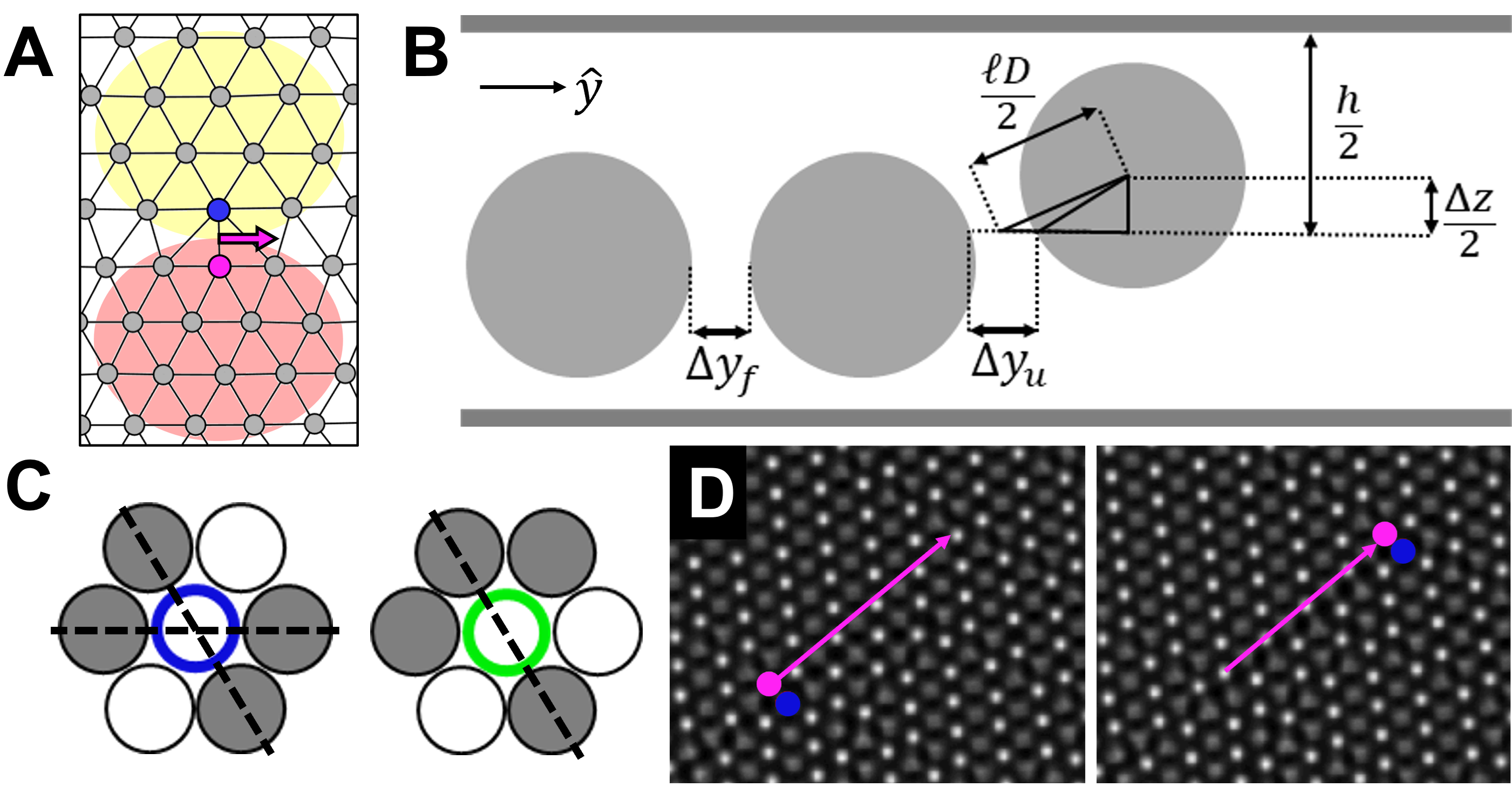}
  \caption{Anisotropic compressibility in stripe and zig-zag spin domains. (a) Lattice dislocations cause compression (red shading) and expansion (yellow shading) on the sides of the particle with five nearest neighbors (magenta) and seven nearest neighbors (blue), respectively. The magenta arrow with black outline indicates the Burgers vector of the lattice dislocation shown. (b) The mean free length between a pair of adjacent same-spin particles is $\Delta y_\mathrm{f} = D(\ell-1)$, while the lateral separation between opposite-spin particles is $\Delta y_\mathrm{u} =   \sqrt{\ell^2 D^2 - \Delta z^2 } - \sqrt{D ^2 - \Delta z^2}$. This additional free distance gives alternating spin rows more room to be compressed along the lattice direction. Note that the looseness $\ell$ is exaggerated here for visual clarity. (c) A spin domain of stripe motifs (left) has one same-spin lattice axis and two alternating spin lattice axes (dotted lines). A spin domain of zig-zag motifs has two half-frustrated axes, where half of the edges are same-spin and half are opposite-spin, and one alternating spin axis (dotted line). (d) An experimentally observed lattice dislocation glides within a spin domain so that the five (magenta) moves along an alternating-spin segment of a lattice line. The glide region is bounded by two pairs of adjacent up spins at the lower left and upper right, confining lattice dislocation motion to a segment of length $x=8$ lattice constants.} 
  \label{fig:2}
\end{figure}

It is well-established that the crystal is compressed on the side of the lattice dislocation nearest to the particle with five neighbors, and expanded on the side of the seven-fold particle \cite{HirthLothe1982}. This means that each particle on the side of the five-fold particle has less free volume surrounding it, while the particles on the seven-fold particle side have more free volume, as depicted in Fig.~\ref{fig:2}(a). For a lattice dislocation to move along its glide axis (the axis parallel to its Burgers vector), there must be sufficient free space on the side of the five-fold particle for the lattice to compress \cite{Burgers1940}.

In flat monolayer crystals with lattice constant $\ell D$, all three lattice directions are equally compressible, because there is on average an equal amount of free space between adjacent particles along each of the lattice directions. We call this lateral separation the ``free length'' along the lattice direction. In flat monolayer crystals, the mean free length along a lattice direction is $D(\ell-1)$, so on average a row of $N+1$ particles can compress by a total distance of $ND(\ell-1)$. However, in a buckled monolayer, the mean free length depends on whether the particles have opposite or same spin (Fig. \ref{fig:2}(b)). For same-spin, frustrated particle pairs, the mean free length along the lattice direction $\hat{y}$ is $\Delta y_\mathrm{f} = D(\ell-1)$, like in flat monolayer crystals. But in a buckled monolayer of height $h$, the mean free length between opposite-spin, unfrustrated particle pairs is $\Delta y_\mathrm{u} =   \sqrt{\ell^2 D^2 - \Delta z^2 } - \sqrt{D ^2 - \Delta z^2}$, which is longer than $\Delta y_f$ due to the curvature of particles' surfaces, as shown in Fig.~\ref{fig:2}(b).

In a spin domain composed of only stripe motifs, one lattice axis is composed of all same-spin particles, so all edges on that axis are frustrated. The other two lattice lines have alternating spin, so all of their edges are unfrustrated. Consequently, lattice dislocation motion is anisotropic in domains containing only stripe motifs: the lattice dislocations are free to move along the two unfrustrated spin axes, which can be easily compressed to accommodate the lattice dislocation as it passes, but motion along the same-spin axis is suppressed. Since this geometric frustration constrains the motion of lattice dislocations, it also influences crystal deformation and growth processes that rely on the motion of lattice dislocations. This is reminiscent of restricted dislocation motion in crystals of dimer-shaped colloidal particles \cite{Gerbode2008} and hints at the possibility of glassy defect dynamics \cite{Zhou2017}.

\begin{figure}[h!]
 \includegraphics[width=0.48\textwidth]{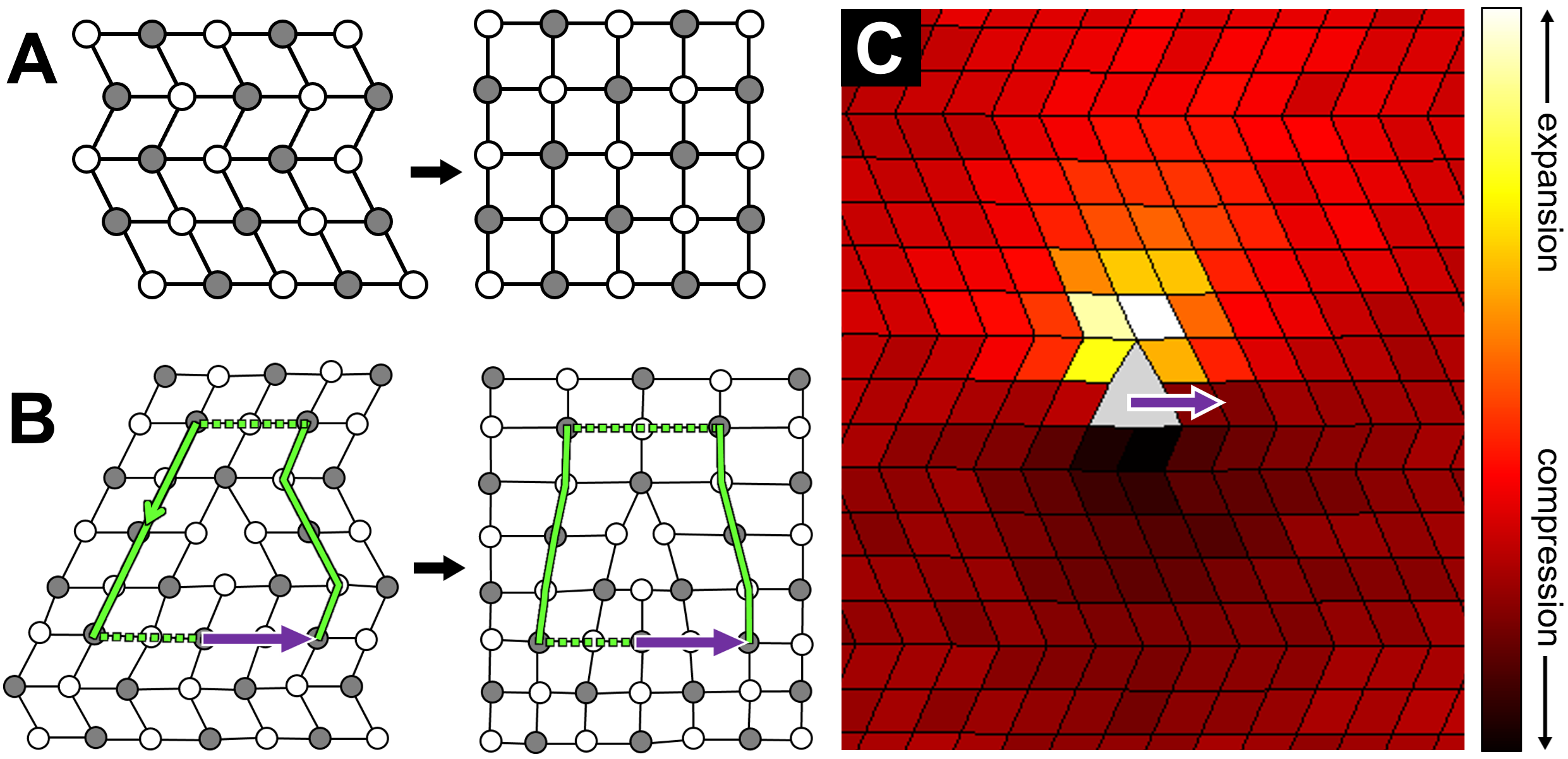}
  \caption{Spin defects correspond to dislocations in the unfrustrated lattice. (a) Any spin domain consisting of combinations of stripe and zig-zag motifs is made up of lines of alternating spin particles. The unfrustrated lattice contains all particle $x$-$y$ positions (white dots - spin up, grey dots - spin down), connected only by opposite-spin nearest-neighbor edges  (black segments). The unfrustrated lattice can be continuously deformed into a square lattice. (b) The presence of a spin defect (in this case a ``pitchfork'' defect) corresponds to a double dislocation in the unfrustrated lattice, which has the effect of adding two lattice rows, as illustrated by the Burgers circuit (green) and the Burgers vector (purple with white outline). (c) Dipole compression-extension field surrounding a simulated pitchfork spin defect. Plaquettes in the unfrustrated lattice are colored linearly with their area, with smaller area in darker color and larger area in brighter color. } 
  \label{fig:3}
\end{figure}

\begin{figure*}[ht!]
 \includegraphics[width=0.98\textwidth]{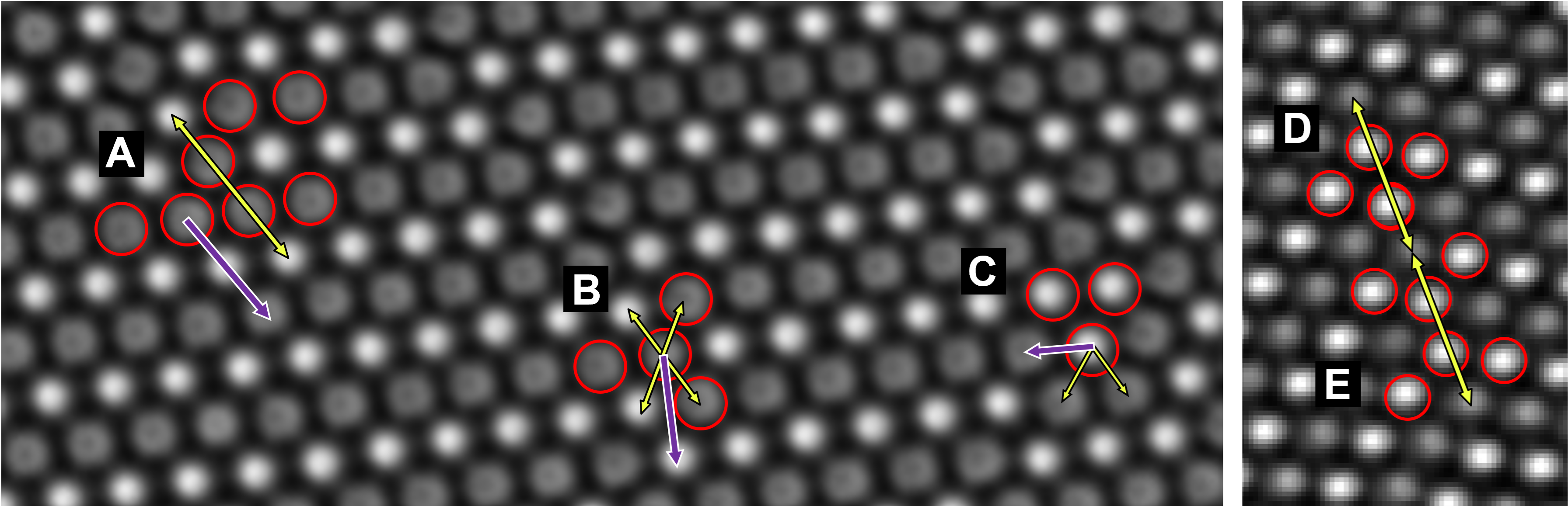}
  \caption{Experimental images containing the five basic spin defects that can be isolated in a spin domain: pitchfork (a), flower~(b), bump (c), diamond (d), and antidiamond (e). Purple arrows (white outlines) indicate the Burgers vector, calculated using a Burgers circuit as shown in Fig. 3. Yellow arrows (black outlines) indicate the direction in which each defect can move due to spin flips along a single alternating spin axis. Note that only the pitchfork has its Burgers vector aligned with its direction of motion, and that the diamond and antidiamond have zero Burgers vector.} 
  \label{fig:4}
\end{figure*}

\section{Spin defects}
\label{spindefects}

A spin domain, whether composed of purely stripe motifs, purely zig-zag motifs, or a combination of the two, is made up of parallel lattice lines of alternating, unfrustrated spins, as shown in yellow in Fig.~\ref{fig:1}(b). We define the ``unfrustrated lattice'' as the lattice consisting of all particle $x$-$y$ coordinates, connected only by nearest-neighbor edges between opposite-spin particles. A spin defect occurs when one of these unfrustrated lattice lines is interrupted by a pair of same-spin particles -- we call the location of the pair of same-spin particles the ``core'' of the spin defect. Five basic spin defects can exist in isolation in the interior of a spin domain. These defects, which we call pitchfork, bump, flower, diamond and antidiamond, are shown in Fig.~\ref{fig:4}. 

\subsection{Topological charge}
 Just as lattice dislocations break the translational order of the triangular lattice in a flat monolayer crystal, spin defects break the translational order of the unfrustrated lattice: the pitchfork, bump and flower defects are dislocations in the unfrustrated lattice, while the diamond and antidiamond are vacancies and interstitials, respectively. To understand the topological charge of each defect, we first observe that the unfrustrated lattice of every spin domain is homeomorphic to a square lattice under a ``brochure''-type transformation, as shown in Fig.~\ref{fig:3}(a). A pitchfork spin defect corresponds to a dislocation in the equivalent square lattice, as shown in Fig.~\ref{fig:3}(b) where we diagram its brochure transformation.

To calculate the topological charge of a spin defect, we use a Burgers circuit to find the Burgers vector \cite{Burgers1940}. Typically, Burgers circuits are taken along straight lattice lines. However, the lattice lines of the unfrustrated lattice change directions at zigzag motifs, so it is not so obvious which edges correspond to which lattice line. But because the brochure transformation is a homeomorphism, we are guaranteed that the straight lattice lines of the square lattice are in bijective correspondence with the bent lattice lines of the unfrustrated lattice. This allows us to identify each Burgers circuit in the unfrustrated lattice with a Burgers circuit from the square lattice (Fig.~\ref{fig:3}(b)). 

The analogy between dislocations in the $x$-$y$ lattice and spin defects extends beyond the topological charge. Importantly, the presence of a pitchfork spin defect causes a dipole compression-extension strain field in the $x$-$y$ lattice, as depicted in Fig.~\ref{fig:3}(c). This compression-extension field was calculated by initializing an isolated pitchfork defect and averaging particle positions over time using Brownian dynamics simulations, which are described in section \ref{sec:coarsening} below.

By transforming the unfrustrated lattice to a square lattice, we determine the Burgers vector for each spin defect -- these are shown in Fig.~\ref{fig:4} as purple arrows with white outlines. Notably, the diamond and antidiamond spin defects have zero Burgers vector, since these correspond to vacancies and interstitials in the homeomorphic square lattice, respectively.

\subsection{Spin defect motion}
\label{sec:motion}
In the bulk of a spin domain, a spin defect moves due to a spin flip at the core of the defect. When one particle at the core of a defect flips spin, the spin defect translates in the direction of the particle that flipped. Spin defects that translate via spin flips along a single alternating axis can move without producing additional defects. Such defects are called ``glissile'' \cite{HirthLothe1982}. In contrast, spin defects that can translate along either of two different directions necessarily absorb or emit other spin defects as they move -- these are ``sessile.'' Since glissile spin defects can glide without producing additional spin defects, they move at no overall cost to the free energy and therefore are more mobile than sessile spin defects.

\subsubsection{Glissile spin defect motion}

The pitchfork, diamond, and antidiamond defects are all glissile, so they have one glide axis. The motion of a glissile spin defect along its glide axis is equivalent to a sequence of consecutive spin flips between the initial and final position of its core. For example, in Fig.~\ref{fig:5}(a,b), a pitchfork defect glides by two unit steps $2 \Delta$ along its axis of motion. The two spins outlined in green between the initial and final position of the spin defect have flipped. We call this motion ``glide'' in analogy with the glide of lattice dislocations, since their direction of motion is aligned with the axis of their Burgers vectors, as shown by the yellow and purple arrows in Fig.~\ref{fig:4}.

The diamond and antidiamond defects move according to the same process as the pitchfork defect in Fig~\ref{fig:5}(a,b). These defects have zero Burgers vector and respectively correspond to a vacancy and an interstitial in the associated homeomorphic square lattice. However, unlike a vacancy or interstitial in a square lattice, which can diffuse in any lattice direction, a diamond or antidiamond defect has only one direction of motion. 

\subsubsection{Sessile spin defect motion}

The bump and flower defects are sessile. A sessile spin defect cannot glide without locally increasing the free energy, but it can move in either of two lattice directions by emitting or absorbing other defects. We call this motion ``climb'' since the axis of motion is not aligned with the Burgers vector. These defects climb by emitting or absorbing a glissile spin defect, similarly to how lattice dislocations in flat monolayer crystals emit or absorb vacancies during climb \cite{HirthLothe1982,McLean1971,Lothe1960,Mott1951}.

\begin{figure}[h!] \includegraphics[width=0.48\textwidth]{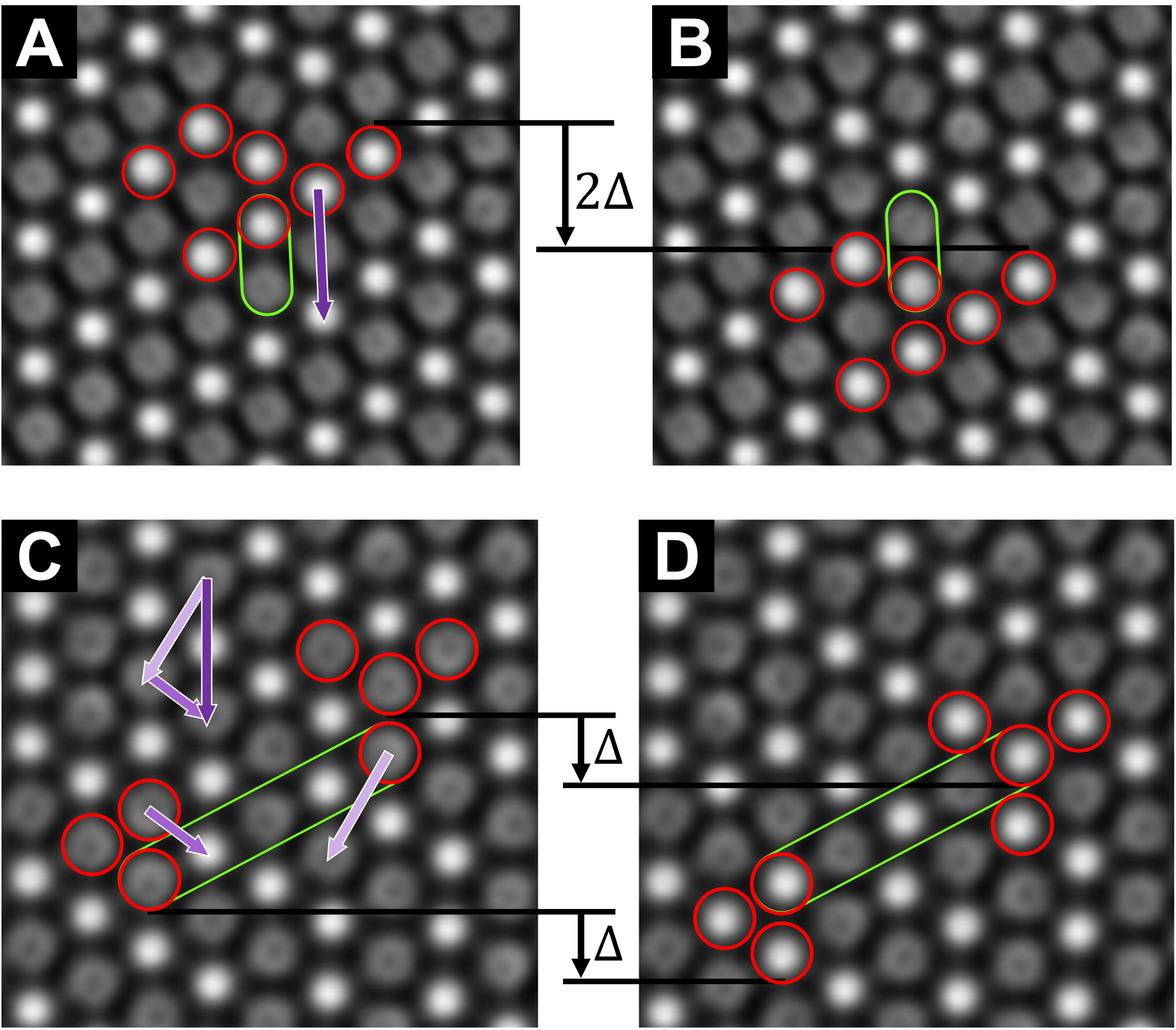}
  \caption{Spin defect motion. (a)-(b) Pitchfork defect glide by two steps ($2\Delta$) via two spin flips (outlined in green). Displacement is along the direction of the Burgers vector (purple arrow, white outline). (c)-(d) Bump and flower defects cannot move independently without emitting additional spin defects. Here a bump and flower defect move together by one step ($\Delta$) via five spin flips (outlined in green). Their displacement is along the direction of the vector sum of their Burgers vectors (dark purple arrow, white outline).}  
  \label{fig:5}
\end{figure}

Although sessile defects cannot glide on their own, combinations of sessile defects whose Burgers vectors combine to that of a glissile defect can glide together. We call this process ``sessile pair glide," and it is mediated by a glissile spin defect. As shown in Fig.~\ref{fig:5}(c,d), two sessile defects translate parallel to the vector sum of their Burgers vectors due to a line of spin flips between them. The overall process produces no additional spin defects because the glissile defect that is emitted at one sessile defect is absorbed at the other.

The glissile defect-mediated translation of sessile defects suggests an interesting idea that warrants further exploration in a future study: the mean free path between sessile defects may set an upper bound on the expected distance that glissile spin defects can travel. It would be interesting to explore whether the evolution of spin domains is dominated by different spin motion mechanisms at different time scales. On short timescales on the order of individual spin flips, the evolution of spin domains may be dominated by the glide of glissile defects, while on longer timescales it may be dominated by the slower motion of the sessile defects.

\subsubsection{Spin defect annihilation}
 Like $x$-$y$ lattice dislocations, spin defects with opposite Burgers vectors can combine and annihilate. For example, two oppositely oriented pitchfork defects can move together and annihilate, conserving Burgers vector during the reaction. We can understand this process from the lens of compression fields such as the one shown in Fig.~\ref{fig:3}(c). In fact, this process is analogous to the annihilation of two lattice dislocations in a triangular crystal. Each pitchfork defect produces a compression field that distorts the unfrustrated lattice, but if these fields are antialigned, then as the two spin defects approach eachother, their two compression fields combine, minimizing the distortion of the lattice. When the two defects annihilate, the distortion is eliminated.

\section{Interactions between spin defects and lattice dislocations}

Both spin defects and lattice dislocations are active in our colloidal experiments. Since both types of defects produce strain fields, which can mutually interact, an effective attraction or repulsion can occur between the different types of defects. We have observed spin defects and lattice dislocations in close proximity both within the bulk of spin domains and at their edges.

\begin{figure}[h!]
 \includegraphics[width=0.5\textwidth]{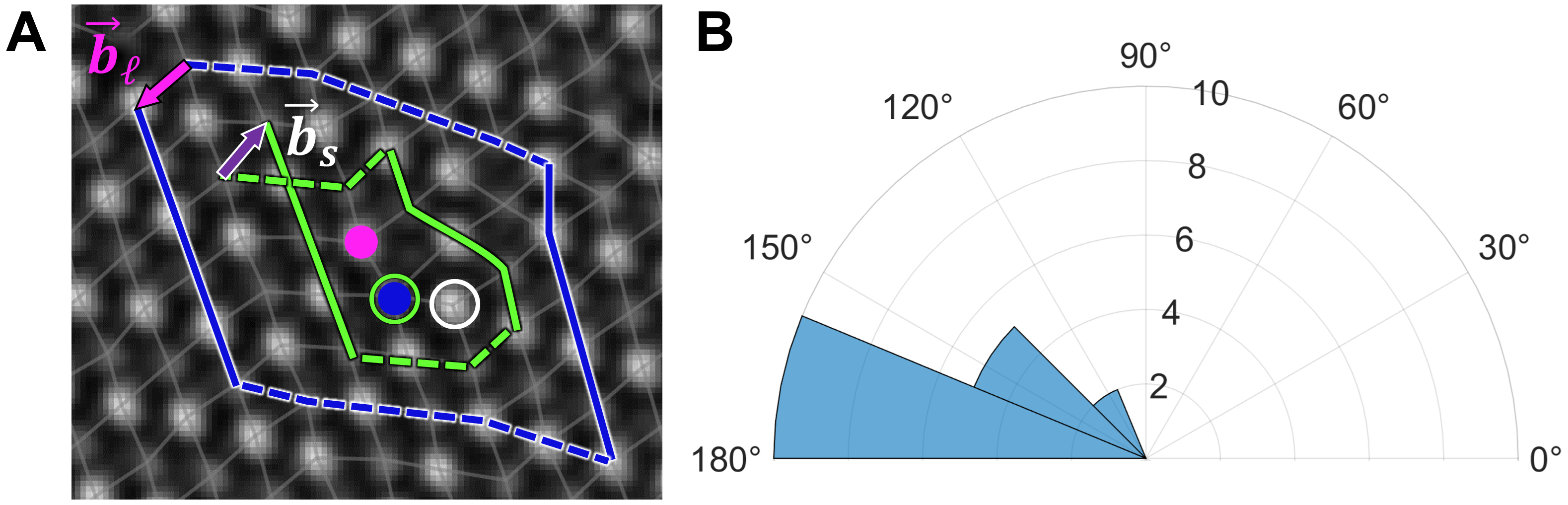}
  \caption{Coupling between spin defects and lattice dislocations within spin domains. (a) A spin defect's core (white and green open circles) and a lattice dislocation's core (magenta and blue dots) are coincident. The Burgers circuit and resulting Burgers vector for the lattice dislocation are shown in blue and magenta outlined in black, respectively. The Burgers circuit and Burgers vector for the spin defect are shown in green and purple outlined in white, respectively. The angle between these Burgers vectors is approximately $180^\circ$. The Burgers circuit for the lattice dislocation steps along nearest neighbor edges, while the Burgers circuit for the spin defect only steps along unfrustrated edges (gray lines).  (b) Polar histogram of measured angles between the Burgers vectors of spin defects and lattice dislocations with coincident cores located within experimentally measured spin domains.}
  \label{fig:interactions}
\end{figure}

\subsection{Coupling between spin defects and lattice dislocations in the bulk}
Lattice dislocations and spin defects interact through their compression fields, which are oriented according to their Burgers vectors. Although the lattice translational order and the frustrated translational order are related, lattice-Burgers vectors (lattice-BV) and spin-Burgers vectors (spin-BV) are independent. For example, there are lattice dislocations with 0 spin-BV and spin defects with 0 lattice-BV. Thus, the Burgers vectors of these two translational orders are independent but can couple to each other according to the orientation of their strain fields.

Since lattice dislocation cores contain one particle with 5 neighbors and another with 7 neighbors, these particles are in neither stripe nor zig-zag motifs, both of which require 6 nearest neighbors. Therefore either of the two particles at the core of a lattice dislocation could simultaneously be at the core of a spin defect. In this case the coincident-core lattice dislocation and spin defect has both a lattice-BV and a spin-BV, as shown in Fig.~\ref{fig:interactions}(a). Isolated lattice dislocations on the interior of a spin domain are rare. Among a sample of $N=17$ isolated lattice dislocations with coincident lattice-BVs and spin-BVs, the distribution of angles between the spin-BV and the lattice-BV is peaked near $180^\circ$, as shown in Fig.~\ref{fig:interactions}(b). This suggests that isolated spin defect/lattice dislocation bound pairs tend to have antiparallel Burgers vectors.

This is similar to the attraction that occurs between lattice dislocations with opposite lattice-BVs. Such lattice dislocations mutually attract and annihilate when they coincide, reducing the overall strain field to zero \cite{HirthLothe1982}. But because the lattice-BV and the spin-BV are associated with independent Burgers circuits, they are not guaranteed to fully annihilate with each other when they coincide. The overall strain field of such a coincident-core pair of lattice/spin defects may be shielded, but does not necessarily vanish. 

\subsection{Coupling at grain boundaries}
Spin defects tend to be closely clustered along grain boundaries, making it challenging to calculate their Burgers vectors. We can only isolate spin defects along low-misorientation grain boundaries, where lattice dislocations are spaced by a larger distance. 

\begin{figure}[h!]
 \includegraphics[width=0.5\textwidth]{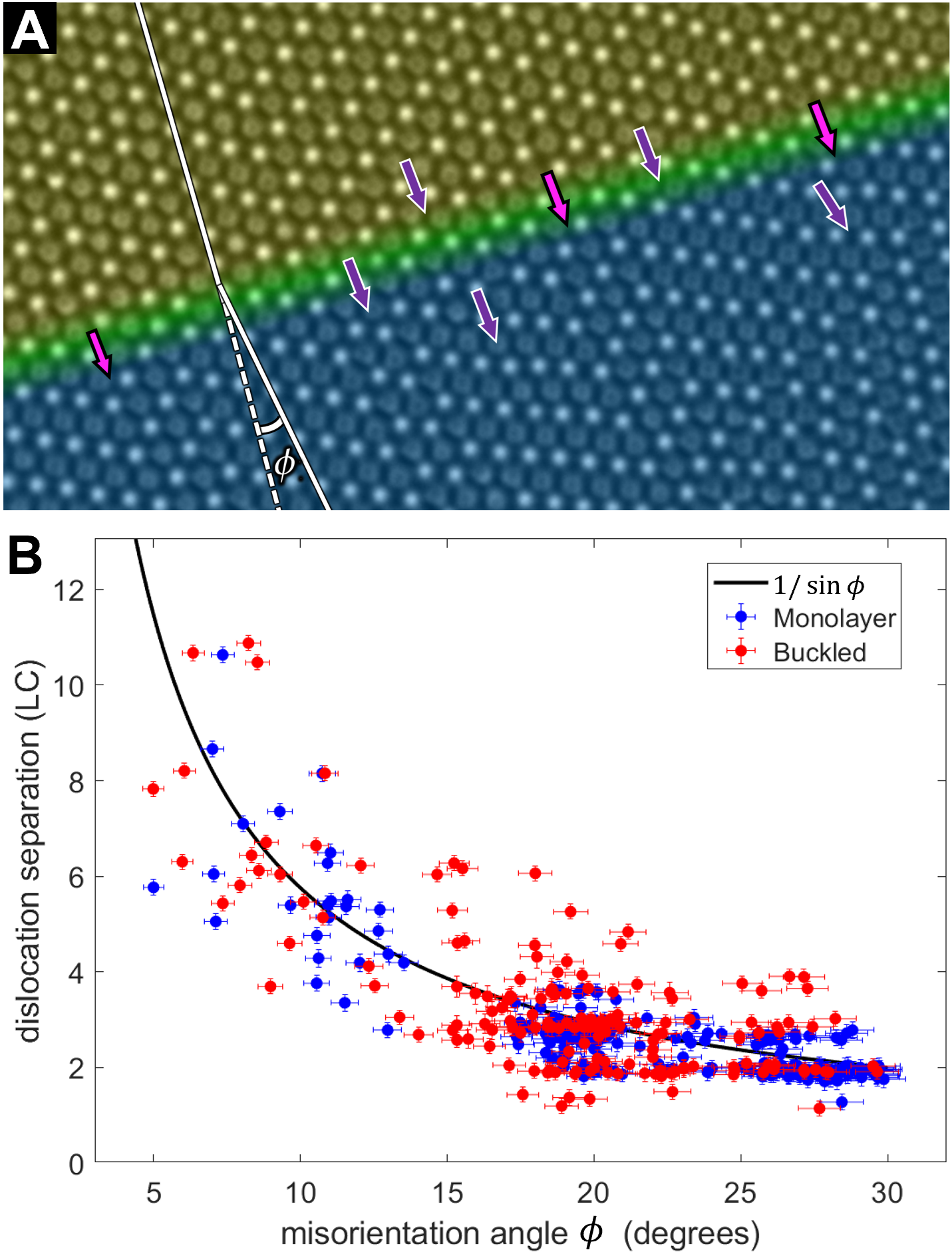}
  \caption{Coupling between spin defects and lattice dislocations at grain boundaries. (a) A low misorientation grain boundary (green shading) separates two spin domains (blue and yellow shading) with a misorientaiton angle $\phi$. The grain boundary is composed of a sequence of lattice dislocations (Burgers vectors: magenta arrows with black outlines) and spin defects (Burgers vectors: purple arrows with white outlines). The orientations of their compression fields align, so that both types of defects contribute to the bending of lattice lines in the grain boundary. (b) The separation (in lattice constants LC) between lattice dislocations in grain boundaries in the buckled monolayer (red) and the flat monolayer (blue) is plotted against the misorientation angle $\phi$. Solid line: theoretical prediction $1/\sin{\phi}$. Separation in the buckled monolayer deviates from the prediction more on average than separation in the monolayer.}
  \label{fig:GB}
\end{figure}

The theory of dislocations predicts that the mean distance between lattice dislocations along a grain boundary with misorientation angle $\phi$ is $1/\sin{\phi}$ lattice constants \cite{Burgers1940}. In some low-misorientation angle grain boundaries, such as the one shown in Fig.~\ref{fig:GB}(a), we observe a sequence of lattice dislocations and spin defects with parallel lattice-BVs and spin-BVs. This suggests that, just like lattice dislocations, spin defects can contribute to the lattice-bending strain fields that occur at grain boundaries.

To investigate the extent of this contribution, we measure the mean distance between lattice dislocations in grain boundaries as a function of $\phi$, and we compare this measurement in flat monolayers to the same measurement in buckled monolayers, as shown in Fig.~\ref{fig:GB}(b). We find that the observed spacing between lattice dislocations in grain boundaries in buckled monolayers varies around the theoretical prediction $1/\sin{\phi}$ with a larger variance than the spacing between lattice dislocations in grain boundaries in the flat monolayer. Specifically, for the flat monolayer, the mean squared difference between the experimental measurements and the $1/\sin{\phi}$ prediction is $0.52 \pm 0.16$ square lattice constants, while for the buckled monolayer it is $1.29 \pm 0.19$ square lattice constants. This suggests that the strain fields of the spin defects do contribute to the misorientation of grain boundaries in the buckled monolayer, either assisting or resisting the effect of the lattice dislocations. We expect that spin defects aligned with lattice dislocations along a grain boundary increase the mean distance between lattice dislocations, and spin defects antialigned with lattice dislocations along a grain boundary decrease the mean distance between lattice dislocations. This may have implications for grain boundary coarsening in buckled monolayer crystals.

\section{Spin domain evolution via spin defects and lattice dislocations}
\label{sec:coarsening}
Coarsening describes the formation and long-term behavior of ordered regions in systems like ours. In the colloidal monolayer, grain coarsening is mediated by lattice dislocation motion. In the antiferromagnetic Ising model, spin domain coarsening is accomplished through spin flips alone \cite{Wannier1950, Wannier1973}. Strikingly, in the buckled monolayer, we observe that spin domain coarsening is mediated by both spin defect motion and lattice dislocation motion. We investigate the relative importance of these two mechanisms at different regimes of free height $\Delta z$ and looseness $\ell$ via Brownian dynamics simulations.

Our Brownian dynamics simulations consist of hard spheres sandwiched between two hard plates that are separated by height $h=\ell D+\Delta z$. We use periodic boundary conditions along the $x$ and $y$ directions. To obtain a random initial condition without overlapping particles, the particle positions are initialized randomly with with diameters equal to 30\% of the target diameter. We then increase the diameter slowly over 100 diffusion times to the target diameter. Here ``diffusion time'' refers to the time that it takes for an isolated full-size particle to diffuse a distance equal to its own diameter. Once the target particle diameter is reached in the simulation, we equilibrate for an additional 150 diffusion times. Simulation details are provided in supplemental section S3 \cite{suppmat}. 

\begin{figure}[h!]
 \includegraphics[width=0.5\textwidth]{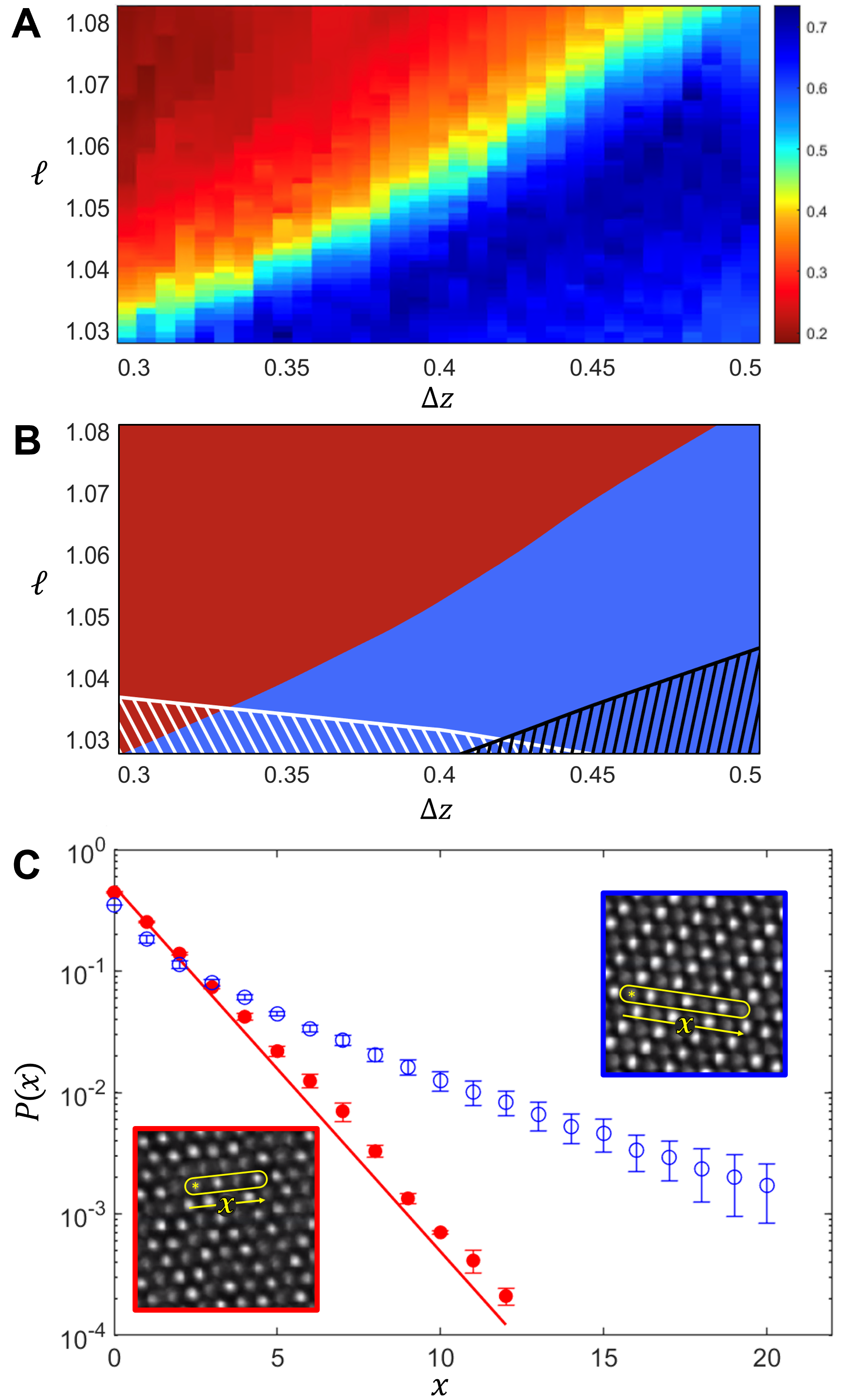}
  \caption{Phase diagram of Brownian dynamics simulations of the buckled colloidal monolayer. (a) \textbf{Phase diagram of spin order}: The fraction of particles in either stripe or zig-zag spin motifs is shown as a heatmap as a function of $\Delta z$ and looseness $\ell$. The phase boundary separates the spin-disordered spin liquid region (red) from the spin solid region, where spins are ordered into spin domains (blue). (b) \textbf{Phase diagram of defect activity}: The red and blue regions represent the spin liquid and spin solid regimes from the phase diagram in (a). In the region with black stripes, spin flips are frozen out, while in the region with white stripes, lattice dislocation motion is frozen out.   (c) \textbf{Lattice dislocation glide distances}: Experimentally measured distribution of number of consecutive alternating spins along each lattice line from every particle for the spin liquid regime (red, solid markers) and spin solid regime (blue, open markers). The spin liquid distribution closely matches the expected result for completely random spins (red line), while the spin solid distribution has a tail indicating preference for alternating spins. This effect helps lattice dislocation motion continue to coarsen spin domains at high $\Delta z$ and low $\ell$.} 
  \label{fig:phaseDiagram}
\end{figure}

We identify regions of the $\Delta z$-$\ell$ phase space where spin domains form by measuring the fraction of particles that are in either stripe or zig-zag motifs following the equilibration period (Fig.~\ref{fig:phaseDiagram}(a)). As shown in Fig.~\ref{fig:phaseDiagram}(a) in the high $\Delta z$, low $\ell$ region, the system is in a spin solid state with large spin domains. Here spins rarely flip because the set of immediately accessible positions to a typical particle does not span from the lower to upper plate. For example, a typical ``up'' particle can access only a set of positions that is clustered near the upper plate. On the other hand, in the low $\Delta z$, high $\ell$ region of Fig.~\ref{fig:phaseDiagram}(a), the system is in a spin liquid state \cite{Balents2010} with ordered $x$-$y$ positions but disordered spins. Here the spins are free to flip because for a typical particle, the set of immediately accessible positions \emph{does} span from the lower to upper plate.

We seek to determine which defect mechanisms are activated in each region of the $\Delta z$-$\ell$ phase space. To determine where spin defects are active, we measure the occurrence of spin flips, since each spin flip causes a glissile spin defect to translate by one lattice constant, as discussed in Section~\ref{spindefects}. To find lattice dislocation activity, we measure the changes to the nearest neighbor edges in the graph of $x$-$y$ positions, since the translation of a lattice dislocation eliminates old nearest neighbor edges and creates new ones. The background of Fig.~\ref{fig:phaseDiagram}(b) is partitioned into red and blue regions to respectively represent the spin liquid and spin solid regions obtained via Fig.~\ref{fig:phaseDiagram}(a). On this background, we highlight with black stripes regions where spin flips are frozen out, and with white stripes regions where lattice dislocation motion is frozen out, conditions that we define as follows: After equilibration, simulated buckled monolayers are observed for a time interval of 50 particle diffusion times. Spins are considered frozen out if fewer than 3\% of all spins flipped, and lattice dislocations are frozen out if fewer than 0.3\% of nearest-neighbor edges in the $x$-$y$ lattice changed, as described in supplemental section S3(B) \cite{suppmat}.

In the region where lattice dislocations cannot move but spin defects can (white stripes only), the particles are so tightly packed in the $x$-$y$ plane that lattice dislocations cannot move. Here only spin defects are active, and these defects dominate spin domain coarsening. In the region where spin defects cannot move but lattice dislocations can (black stripes only), spin flips are suppressed because the dramatic buckling caused by the larger $\Delta z$ increases the amount of overlap between adjacent particles when projected on the $x$-$y$ plane. Thus, neighboring particles need to rearrange further in order for a particle to flip its spin, but the lower looseness inhibits this rearrangement. Yet in this regime, spin domain coarsening still occurs via rearrangements in the $x$-$y$ lattice, which are effected by the motion of lattice dislocations.

We observe that the region in which spin flips are frozen out and only lattice dislocations can move (black stripes) corresponds to the formation of large spin domains. It is interesting to consider how lattice dislocations alone can drive spin domain coarsening. To understand this, we recall the observation that lattice dislocations tend to move along alternating lattice lines in the buckled monolayer, as discussed in Section~\ref{dislocations}. Thus, longer segments of unfrustrated spin between two spin defect cores along a lattice line enable longer lattice dislocation glide. Fig.~\ref{fig:phaseDiagram}(c) shows the distribution $P(x)$ of experimentally-measured distances $x$ from any lattice site to the nearest spin defect, measured along a lattice direction. This distribution was measured for a spin liquid buckled monolayer (red series) and a spin solid buckled monolayer (blue series). Data at longer $x$ corresponds to longer segments of alternating spins that are sandwiched between two spin defect cores. Thus $P(x)$ estimates the distribution of glide lengths accessible to a free lattice dislocation with random position and Burgers vector.

In the spin liquid region, where $\ell$ is large, there is sufficient space between even same-spin particles to accommodate lattice dislocation motion, even through spin defects. In the spin solid region, where $\ell$ is smaller, lattice dislocation motion is restricted, but longer distances between spin defects enable lattice dislocation glide. Accordingly, we have measured that $x$ is longer in this region of the phase diagram. This suggests that lattice dislocations can move further before encountering repeated spins: the lack of spin defects enables sufficient lattice dislocation mobility to effect spin domain coarsening through lattice dislocation motion alone.

Importantly, we find that most of the spin domain region shown in blue in the phase diagram of Figure \ref{fig:phaseDiagram} corresponds to systems where \textit{both} spin defects and lattice dislocations are active. In this region of the phase diagram, both types of topological defects are involved in the growth and coarsening of both spin domains and crystal grains of the $x$-$y$ lattice. Further studies are needed to fully understand how the interactions between spin defects and lattice dislocations influence the dynamics of coarsening, both of the spin domains and of the grain boundaries in the underlying $x$-$y$ lattice, such as in Fig.~\ref{fig:GB}(a). 

\section{Conclusions}

The development of the theory of crystal defects in the 1920's transformed our ability to understand and predict the behavior of crystalline materials. Here we have presented a description of the fundamental defects of both the $x$-$y$ lattice and particle spin, or $z$ position, in a simple colloidal crystal system exhibiting geometric frustration. The additional degree of freedom arising from excess vertical space not only affects the motions of dislocations in the underlying $x$-$y$ lattice but also enables the presence of a new set of topological defects in the spin order. We find that both lattice dislocations and spin defects play a role in the evolution of spin order and $x$-$y$ lattice order in the system, and have determined under what conditions each type of process dominates. Ultimately these results lay the groundwork for modeling fundamental defect mechanisms in a simple system with geometric frustration. Interestingly, interactions between the two types of defects can affect their motion, which is reminiscent of other systems with restricted defect motion that exhibit glassy behavior \cite{Gerbode2008, Gerbode2010}. Since the Burgers vectors and resulting strain fields of both types of topological defects vary throughout the phase diagram, the interactions between these defects and their impact on coarsening will also vary. Future studies may elucidate how the interplay between spin order and order in the $x$-$y$ lattice affects the material properties of this system as well as its ordering and melting mechanisms.

\bibliography{refs}
\end{document}